## CP VIOLATION IN THE B SYSTEM:
## WHAT'S OLD, WHAT'S NEW?[1]

David London

Laboratoire de physique nucléaire, Université de Montréal
C.P. 6128, succ. centre-ville, Montréal, QC, Canada H3C 3J7

**Abstract**

I review the many ways of obtaining information about the CKM matrix through CP violation in the $B$ system. I include direct and indirect CP violation, the role of penguins and isospin analysis, tagging, and $B \to DK$ decays. I also discuss recent developments showing how to use $SU(3)$ flavor symmetry, along with some dynamical approximations, to extract from $B$ decays to $\pi\pi$, $\pi K$ and $K\overline{K^0}$ the weak CKM phases, the strong phase shifts, and the sizes of the various contributing diagrams. Finally, I briefly show what can be learned from an exact treatment of the CKM matrix.

## INTRODUCTION

In the standard model (SM), CP violation is due to the presence of a complex phase in the Cabibbo-Kobayashi-Maskawa (CKM) matrix [1]. The goal of the study of CP violation in the $B$ system is to test this picture of CP violation. In this talk I will review the current wisdom regarding CP violation in $B$ decays (most of which can be described in terms of triangles), as well as some recent developments (more triangles).

When studying CP violation in the $B$ system it is useful to use an approximate form of the CKM matrix due to Wolfenstein [2], which incorporates the experimental fact that the elements of the CKM matrix obey a hierarchy in terms of the Cabibbo angle, $\lambda = 0.22$:

$$\begin{pmatrix} V_{ud} & V_{us} & V_{ub} \\ V_{cd} & V_{cs} & V_{cb} \\ V_{td} & V_{ts} & V_{tb} \end{pmatrix} \sim \begin{pmatrix} 1 - \frac{1}{2}\lambda^2 & \lambda & |V_{ub}|\exp(-i\gamma) \\ -\lambda & 1 - \frac{1}{2}\lambda^2 & A\lambda^2 \\ |V_{td}|\exp(-i\beta) & -A\lambda^2 & 1 \end{pmatrix} . \qquad (1)$$

Here, $A$ is a parameter of $O(1)$, $|V_{ub}|$ and $|V_{td}|$ are terms of $O(\lambda^3)$, and $|V_{ub}|$, $|V_{td}|$, $\beta$ and $\gamma$ are all functions of a single complex phase and one additional real parameter. In this approximation, the only non-negligible phases ($\beta$ and $\gamma$) appear in the terms $V_{ub}$ and $V_{td}$.

Unitarity of the CKM matrix implies, among other things, the orthogonality of the first and third columns:

$$V_{ud}V_{ub}^* + V_{cd}V_{cb}^* + V_{td}V_{tb}^* = 0 . \qquad (2)$$

---

[1] Talk given at the $5^{th}$ *Conference on the Intersections of Particle and Nuclear Physics*, St. Petersburg, Florida, USA, May 31-June 6, 1994.

hep-ph/9406412  28 Jun 1994

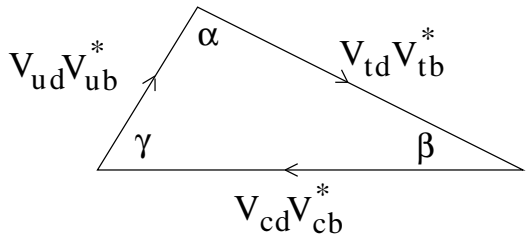

Figure 1: The unitarity triangle.

This relation can be represented as a triangle in the complex plane (Fig. 1), which has come to be known as the *unitarity triangle*. In the Wolfenstein approximation, the angles in the unitarity triangle are given by $\beta = -\text{Arg}(V_{td})$, $\gamma = \text{Arg}(V_{ub}^*)$, and $\alpha = \pi - \beta - \gamma$ [3]. The goal then is to independently measure the three angles of the unitarity triangle, $\alpha$, $\beta$ and $\gamma$, to see (i) if they are all nonzero, and (ii) if they add up to 180°. This will be the acid test for the SM picture of CP violation. In the following sections I will describe the many ways of getting at these three CKM angles through CP violation in $B$ decays.

## WHAT'S OLD?

### CP-violating rate asymmetries

The easiest way to see CP violation in the $B$ system is to look for a rate asymmetry in the decay of a $B$ meson to some final state $f$, i.e. $\Gamma(B \to f) \neq \Gamma(\bar{B} \to \bar{f})$. In order to produce such an asymmetry, it is necessary that there be two interfering weak amplitudes in the process $B \to f$. This can come about in two distinct ways, called *direct* and *indirect (mixing-induced)* CP violation. I will discuss these two in turn.

● *Direct CP Violation*

In order to produce direct CP violation, one needs to have two amplitudes contributing to the decay of a $B$ meson. For example, consider the decay $B^+ \to \pi^0 K^+$. There are two diagrams (tree and penguin) which lead to this decay (Fig. 2). To each diagram is associated a 'weak' and a 'strong' phase. The weak phases $\phi$ come from the CKM matrix, and change sign in going from the decay to the CP-conjugate decay. By contrast, the strong phases $\delta$ are the same for both the decay and the CP-conjugate decay. This is because such phases are due to strong-interaction rescattering and hadronization, and QCD is sensitive to colour only – it is irrelevant whether a quark or an antiquark is involved. The amplitudes for $B^+ \to \pi^0 K^+$ and $B^- \to \pi^0 K^-$ can thus be written

$$
\begin{aligned}
A(B^+ \to \pi^0 K^+) &= T\, e^{i\phi_T}\, e^{i\delta_T} + P\, e^{i\phi_P}\, e^{i\delta_P} \,, \\
A(B^- \to \pi^0 K^-) &= T\, e^{-i\phi_T}\, e^{i\delta_T} + P\, e^{-i\phi_P}\, e^{i\delta_P} \,.
\end{aligned}
\tag{3}
$$

It is straightforward to calculate the rate asymmetry. It is

$$
\Gamma(B^+ \to \pi^0 K^+) - \Gamma(B^- \to \pi^0 K^-) \sim \sin(\phi_T - \phi_P)\sin(\delta_T - \delta_P).
\tag{4}
$$

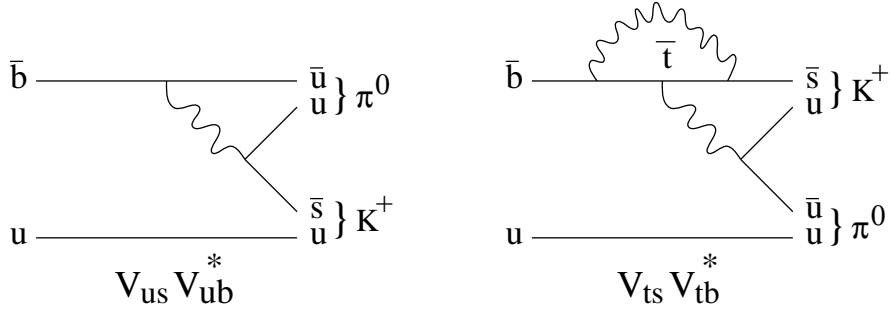

Figure 2: Two diagrams contributing to the process $B^+ \to \pi^0 K^+$.

One sees that, in order to have direct CP violation, it is necessary to have both a weak phase difference and a strong phase difference between the contributing diagrams. In this particular case, $\sin(\phi_T - \phi_P) \sim \sin\gamma$. Unfortunately, however, the strong phase differences are incalculable. Thus, even if a CP-violating rate asymmetry were measured in this mode, it would not give *clean* information regarding the CKM phase $\gamma$, since we cannot disentangle it from the strong phases. This is a generic feature of direct CP violation – unless special techniques are used (which I will describe below), one cannot cleanly extract information about the weak CKM phases. (In this particular case, flavour $SU(3)$ can be used – I will come back to this decay later.)

- *Indirect (Mixing-Induced) CP Violation*

For indirect CP violation, one uses rate asymmetries which are due to $B^0$-$\overline{B^0}$ mixing. That is, one considers a final state $f$ to which both $B^0$ and $\overline{B^0}$ can decay. CP violation then comes about via the interference of the two amplitudes $B^0 \to f$ and $B^0 \to \overline{B^0} \to f$. The beauty of indirect CP violation is that it is possible to obtain clean information about the weak CKM phases. In order to do so, it is a necessary requirement that only one weak amplitude contribute to the decay. If more than one amplitude contributes, then direct CP violation is introduced, ruining the cleanliness of the measurement.

For this type of analysis it is necessary to measure the four time-dependent rates $B^0(t) \to f$, $\overline{B^0}(t) \to f$, $B^0(t) \to \bar{f}$ and $\overline{B^0}(t) \to \bar{f}$ [4]. Here, $B^0(t)$ [$\overline{B^0}(t)$] is a state which is produced as a $B^0$ [$\overline{B^0}$] at time $t = 0$. Due to $B^0$-$\overline{B^0}$ mixing it will evolve in time into a mixture of $B^0$ and $\overline{B^0}$. If $f$ is a CP eigenstate, then the above four rates reduce to two rates.

This type of CP asymmetry can be divided into four classes, three of which are expected to be nonzero in the SM [3]. Along with the CKM angles measured, they are:

1. $\overset{(-)}{B_d}$ decays with $b \to u$ (e.g. $\overset{(-)}{B_d} \to \pi^+\pi^-$): $\sin 2\alpha$

2. $\overset{(-)}{B_d}$ decays with $b \to c$ (e.g. $\overset{(-)}{B_d} \to \Psi K_s$): $\sin 2\beta$

3. $\overset{(-)}{B_s}$ decays with $b \to u$ (e.g. $\overset{(-)}{B_s} \to D_s^+ K^-$, $D_s^- K^+$ [5]): $\sin^2\gamma$

4. $\overset{(-)}{B_s}$ decays with $b \to c$ (e.g. $\overset{(-)}{B_s} \to \Psi\phi$): 0

(Note that the CP asymmetry in this last class of decays is zero only in the Wolfenstein approximation; in fact, this asymmetry is nonzero, albeit small ($\lesssim 0.05$), if the CKM matrix is treated exactly. I will come back to this point later.) As an example, consider the decays $\overset{(-)}{B_d} \to \Psi K_S$. When $B^0$-$\overline{B^0}$ mixing is taken into account, the rates take the following form:

$$
\begin{aligned}
\Gamma\left(B_d^0(t) \to \Psi K_S\right) &= e^{-\Gamma t}\left[1 - \sin 2\beta \sin x_d t\right], \\
\Gamma\left(\overline{B_d^0}(t) \to \Psi K_S\right) &= e^{-\Gamma t}\left[1 + \sin 2\beta \sin x_d t\right].
\end{aligned}
\tag{5}
$$

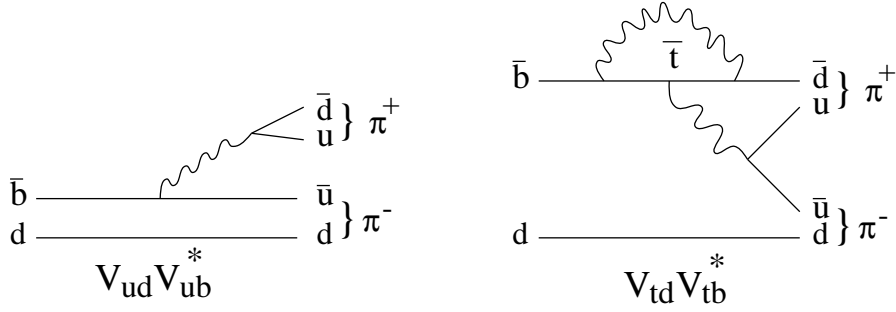

Figure 3: Diagrams contributing to the process $B_d^0 \to \pi^+ \pi^-$.

Here, $x_d = (\Delta M)/\Gamma$ is the $B_d^0$-$\overline{B_d^0}$ mixing parameter. From these equations, one sees quite clearly that the measurement of the rates for $\overset{(-)}{B_d} \to \Psi K_S$ allows the clean extraction of the CKM angle $\beta$. I stress, however, that for measurements of indirect CP violation, it is, in general, necessary (i) to measure the time development, and (ii) to know whether the decaying $B$ was a $B^0$ or $\overline{B^0}$ at $t = 0$ (i.e. *tagging* is needed).

### Penguin pollution and isospin analysis

As mentioned above, a necessary requirement for the clean extraction of CKM phases from measurements of indirect CP violation is that only one weak amplitude contribute to the decay. A problem arises, however, when one realizes that penguin diagrams might play a role in these decays [6]. Such diagrams will contribute to $\overset{(-)}{B_d} \to \pi^+ \pi^-$, for example (Fig. 3). In this case, since the tree and penguin diagrams have different weak phases ($\mathrm{Arg}(V_{ub}^* V_{ud}) \sim \gamma$ and $\mathrm{Arg}(V_{tb}^* V_{td}) \sim -\beta$, respectively), they will interfere, giving rise to direct CP violation in addition to the indirect CP violation. As usual, the presence of direct CP violation spoils the cleanliness of the measurement – the CP-violating asymmetry is no longer proportional to $\sin 2\alpha$, but rather $\sin(2\alpha + \theta_{+-})$, where $\theta_{+-}$ depends on the weak and strong phases of the tree and penguin diagrams, as well as on their relative sizes. The quantity $\theta_{+-}$ is, of course, incalculable. (Note that this problem does not arise for the other two CP asymmetries – there are no penguin contributions to $\overset{(-)}{B_s} \to D_s^+ K^-$, $D_s^- K^+$, and in the case of $\overset{(-)}{B_d} \to \Psi K_S$ the tree and penguin diagrams have the same weak phase, so there is no interference.)

Fortunately, it is still possible to clean up this measurement by using isospin symmetry [7]. Consider the three processes $B_d^0 \to \pi^+ \pi^-$, $B_d^0 \to \pi^0 \pi^0$ and $B^+ \to \pi^+ \pi^0$. The $B$ meson has isospin $1/2$; the two final pions must be in a state of total isospin $I = 0$ or $2$. Since there are two isospin amplitudes ($\Delta I = 1/2, 3/2$), but three processes, there must be a triangle relation. There is a similar relation among the CP-conjugate processes $\overline{B^0_d} \to \pi^+\pi^-$, $\overline{B^0_d} \to \pi^0\pi^0$ and $B^- \to \pi^-\pi^0$. These triangle relations are

$$\frac{1}{\sqrt{2}}A^{+-} + A^{00} = A^{+0} . \tag{6}$$

$$\frac{1}{\sqrt{2}}\bar{A}^{+-} + \bar{A}^{00} = \bar{A}^{-0} . \tag{7}$$

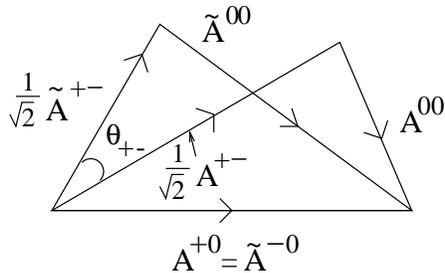

Figure 4: Isospin triangles in $B \to \pi\pi$.

There are two points worth noting. First, the penguin diagram is pure $\Delta I = 1/2$, but the tree diagram has both $\Delta I = 1/2$ and $\Delta I = 3/2$ pieces. Thus, if we could isolate the $\Delta I = 3/2$ component of $B^0_d \to \pi^+\pi^-$, we would have eliminated the penguin pollution. Second, the decay $B^+ \to \pi^+\pi^0$ $(A^{+0})$ is pure $\Delta I = 3/2$ – there is no penguin contribution.

The key point is that, since $A^{+0}$ has one amplitude only, there can be no CP violation in this decay. Thus $|A^{+0}| = |\bar{A}^{-0}|$, so that the two triangles have a base in common. (Since $A^{+-}$ and $A^{00}$ have contributions from both isospin amplitudes, there will in general be direct CP violation, so we expect that $|A^{+-}| \neq |\bar{A}^{+-}|$ and $|A^{00}| \neq |\bar{A}^{00}|$.) The fact that the two triangles have a common base permits the experimental measurement of the incalculable quantity $\theta_{+-}$. By measuring all the decay rates, one can construct the two triangles, as in Fig. 4. (In this figure, the $\bar{A}$'s are related to the $\bar{A}$'s by a rotation.) From this figure we see that, up to a discrete ambiguity (since one or both triangles may be flipped upside down), $\theta_{+-}$ can be determined. With this knowledge the angle $\alpha$ can be extracted by measuring CP violation in $B^0_d(t) \to \pi^+\pi^-$. Thus, even in the presence of penguins, $\alpha$ can be obtained cleanly by using the above isospin analysis.

### Tagging

All clean measurements of the CKM phases discussed so far require tagging, i.e. knowing whether the decaying neutral $B$ meson was a $B^0$ or $\overline{B^0}$ at time $t = 0$. At an $e^+e^-$ collider operating at the $\Upsilon(4s)$, this is relatively easy. The $\Upsilon(4s)$ decays to a $B^0\overline{B^0}$ pair in a coherent $C = 1$ state. This means that one neutral $B$ meson cannot oscillate independently of the other one. Thus, if, for example, the $B^0$ decays to the final state $f$ at $t = 0$, the other neutral $B$ meson has to be a $\overline{B^0}$ at this time. Its subsequent semileptonic decay tells us that it was a $\overline{B^0}$ (i.e. by the charge of the lepton), and hence tags the flavour of the $B^0$. While it is true that the $\overline{B^0}$ can oscillate before decaying, we know how to take this into account. However, in order to do this, we have to know which of the $B^0$ or $\overline{B^0}$ decayed first. In other words, time-dependent measurements are necessary, which requires an asymmetric $e^+e^-$ collider. Tagging cannot be done at a symmetric $e^+e^-$ collider.

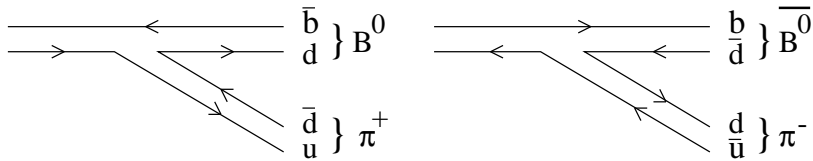

Figure 5: *b*-quark fragmentation diagrams.

At hadron colliders, on the other hand, neutral $B$'s are produced incoherently, i.e. in the states $B^0\overline{B^0}$, $B^0B^-$, $B^0\overline{B_s^0}$, $B^0\Lambda_b$, etc. Some of the partner $B$'s ($\overline{B^0}$, $\overline{B_s^0}$) can undergo mixing, while others ($B^-$, $\Lambda_b$) cannot. Thus, if one wants to use the semileptonic decay to tag the $B^0$, it is necessary to know the production cross sections for each of the above final states. Since these are not well known, this introduces an enormous uncertainty, and makes tagging at hadron colliders somewhat problematic.

A possible solution is to look for correlated $\pi$'s. The idea is that there is a correlation between the flavour of neutral $B$ mesons and the charge of an accompanying pion which makes a low-mass $B$-$\pi$ system [8]. There are two motivations for believing that this might be the case. First, there could exist positive-parity "$B^{**}$" resonances ($J^P = 0^+$, $1^+$, $2^+$, ...). In this case, the $B^{**}$ decays to $B^0\pi^+$ and *not* $B^0\pi^-$. Second, in *b*-quark fragmentation, the leading $\pi$ carries information about the flavour of the neutral $B$ (see Fig. 5) – $B^0\pi^+$ and $\overline{B^0}\pi^-$ tend to be favoured. What is particularly interesting about this idea is that the correlation can be *measured* experimentally. If it turns out to be large, then this will be a very useful technique, and will greatly reduce the tagging error at hadron colliders.

$$B \to DK$$

Now, tagging and time-dependent measurements are *hard*. So, one might ask if it possible to cleanly measure CKM phases without these, i.e. by using rates alone. The answer to this question is *YES*.

One suggestion [9] is to look for a rate asymmetry in the decay $B^+ \to D_{CP}^0 K^+$, where

$$D_{CP}^0 = \frac{1}{\sqrt{2}}(D^0 + \overline{D^0}). \tag{8}$$

That is, $D_{CP}^0$ is a $D^0$ or $\overline{D^0}$ which is identified in a CP-eigenstate mode (e.g. $\pi^+\pi^-$, $K^+K^-$, ...). Since this CP asymmetry involves only charged $B$'s, neither tagging nor time dependence are needed.

Such an asymmetry would be a signal of direct CP violation, which requires two weak decay amplitudes. For this decay, the two contributing amplitudes are simply the individual decays $B^+ \to D^0K^+$ and $B^+ \to \overline{D^0}K^+$, as shown in Fig. 6. These

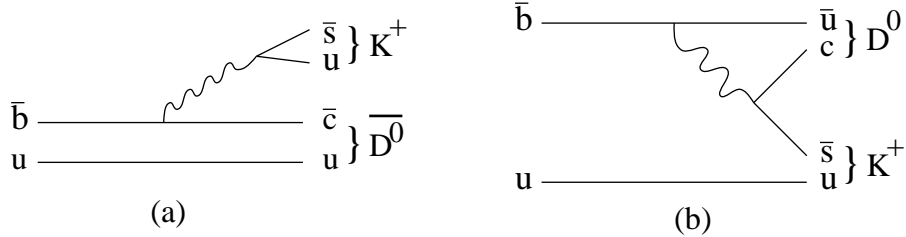

Figure 6: Diagrams contributing to (a) $B^+ \to \overline{D^0}K^+$ and (b) $B^+ \to D^0 K^+$.

amplitudes can be written

$$
\begin{aligned}
A(B^+ \to \overline{D^0}K^+) &= |A_1|\, e^{i\delta_1}\,, \\
A(B^+ \to D^0 K^+) &= |A_2|\, e^{i\gamma}\, e^{i\delta_1}\,,
\end{aligned}
\tag{9}
$$

where $\gamma$ is the weak phase in $B^+ \to D^0 K^+$ and $\delta_{1,2}$ are the strong phases.

Of course, simply measuring the CP asymmetry in $B^+ \to D^0_{CP}K^+$ will not yield the CKM phase $\gamma$ cleanly, since this asymmetry depends also on the difference of the strong phases (c.f. Eq. 4). However, this can be resolved by using the triangle relations which follow from Eq. 8 above:

$$
\begin{aligned}
\sqrt{2}A(B^+ \to D^0_{CP}K^+) &= A(B^+ \to D^0 K^+) + A(B^+ \to \overline{D^0}K^+)\,, \\
\sqrt{2}A(B^- \to D^0_{CP}K^-) &= A(B^- \to \overline{D^0}K^-) + A(B^- \to D^0 K^-)\,.
\end{aligned}
\tag{10}
$$

Let's look more closely at these triangles. There is only one amplitude contributing to each of the individual decays $B^+ \to D^0 K^+$ and $B^+ \to \overline{D^0}K^+$, so that

$$
\begin{aligned}
|A(B^+ \to D^0 K^+)| &= |A(B^- \to \overline{D^0}K^-)|\,, \\
|A(B^+ \to \overline{D^0}K^+)| &= |A(B^- \to D^0 K^-)|\,.
\end{aligned}
\tag{11}
$$

Furthermore, from Eq. 9, there is a relative phase $2\gamma$ between $A(B^+ \to D^0 K^+)$ and $A(B^- \to \overline{D^0}K^-)$. Therefore the two triangles have a common base $[A(B^+ \to \overline{D^0}K^+) = A(B^- \to D^0 K^-)]$, and a second side of the same length. Due to the possibility of CP violation, the third side is not, in general, the same length in both triangles:

$$
|A(B^+ \to D^0_{CP}K^+)| \neq |A(B^- \to D^0_{CP}K^-)|.
\tag{12}
$$

Thus, by measuring 6 time-independent rates only, one can construct the triangles shown in Fig. 7. This shows that the angle $\gamma$ can be extracted cleanly, even though only direct CP violation is involved. There is still a discrete ambiguity due to the possibility of reflection of one of the triangles, corresponding to the exchange $\gamma \leftrightarrow (\delta_1 - \delta_2)$. Also, note that, even if there is no CP violation [i.e. $(\delta_1 = \delta_2)$], the CKM angle $\gamma$ can still be extracted. Since neither tagging nor time-dependence are necessary, this measurement can be done at a symmetric $B$-factory.

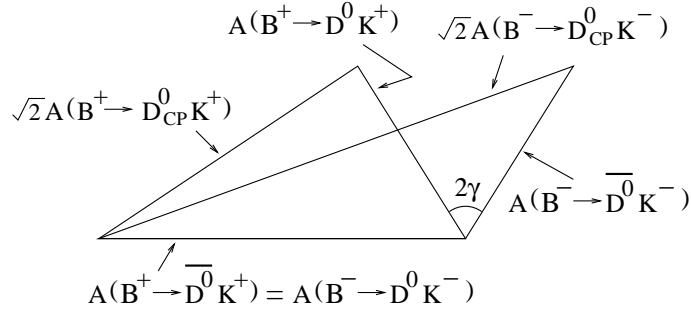

Figure 7: Triangles describing $B \to DK$.

The one possible problem with this method is that $B^+ \to \overline{D^0} K^+$ is colour-allowed, while $B^+ \to D^0 K^+$ is colour-suppressed. Thus, we expect the branching ratios to be $BR(B^+ \to \overline{D^0} K^+) \sim 2 \times 10^{-4}$ and $BR(B^+ \to D^0 K^+) \lesssim O(10^{-5})$. Therefore the triangles are likely to be very thin, which will make the extraction of $\gamma$ more difficult.

## WHAT'S NEW?

### Flavour $SU(3)$

In the previous section I showed that considerations of isospin symmetry can be useful, specifically in distinguishing the weak from the strong phases in the process $B \to \pi\pi$. It is therefore only natural to ask whether flavour $SU(3)$ symmetry can be similarly helpful. After all, this symmetry is expected to hold quite well at the scale of the $b$-quark. In this section I will show that this is indeed the case – it is possible to use $SU(3)$ relations among the amplitudes for $B \to \pi\pi$, $\pi K$ and $KK$ to obtain the weak phases (and more!) cleanly. Furthermore, neither tagging nor time-dependent measurements are needed.

Assuming a flavour $SU(3)$ symmetry [10]-[15], the amplitudes for the decays $B \to \pi\pi$, $\pi K$ and $KK$ can be written in terms of 5 reduced matrix elements. This decomposition is equivalent to an expansion in terms of diagrams. There are 6 diagrams which contribute to the quark-level decay $b \to \bar{u} u \bar{q}$ ($q = d, s$). Shown in Fig. 8, they are: a "tree" amplitude $T$ or $T'$, a "color-suppressed" amplitude $C$ or $C'$, a "penguin" amplitude $P$ or $P'$, an "exchange" amplitude $E$ or $E'$, an "annihilation" amplitude $A$ or $A'$, and a "penguin annihilation" amplitude $PA$ or $PA'$. Here an unprimed amplitude stands for a strangeness-preserving decay, while a primed contribution stands for a strangeness-changing decay. Although there are 6 diagrams, they only ever appear in 5 linear combinations in the $B$-decay amplitudes [10, 13].

Now comes one of the main points. 3 of the 6 diagrams are expected to be smaller than the 3 others for dynamical reasons. More specifically, $E$, $A$ and $PA$ are non-spectator diagrams, that is, they all require that the decaying $b$-quark interact with its partner in the $B$ meson. This leads to a suppression due to the wave function at the origin. Hence $E$, $A$ and $PA$ are expected to be smaller than $T$, $C$ and $P$ by a factor $f_B/m_B \approx 1/25$ (and similarly for their primed counterparts). Therefore, considerations of flavour $SU(3)$ symmetry + dynamics lead to a decomposition of $B$-decay amplitudes in terms of the six diagrams $T$, $T'$, $C$, $C'$, $P$ and $P'$. Furthermore, these diagrams are not all independent – we have $T'/T = C'/C = r_u$ where $r_u = V_{us}/V_{ud} \approx 0.23$. Thus, the amplitudes for the decays $B \to \pi\pi$, $\pi K$ and $KK$ can be written in terms of a very few diagrams, leading to many relations among them.

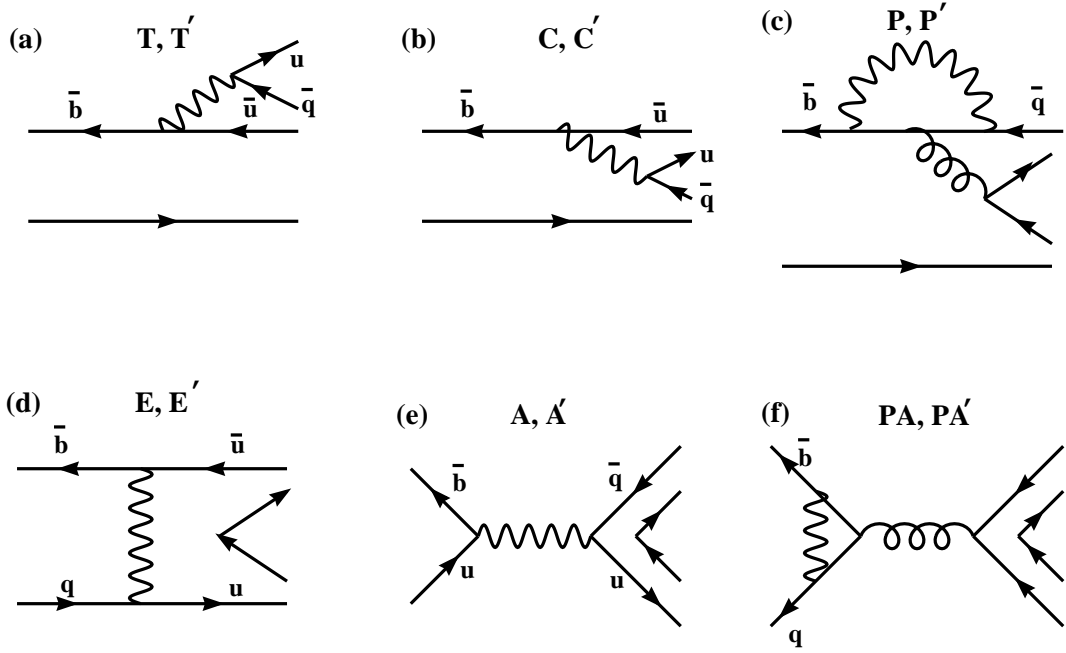

Figure 8: Diagrams describing decays of $B$ mesons to pairs of light pseudoscalar mesons.

As an example, consider the three decays $B^+ \to \pi^+\pi^0$, $B^+ \to \pi^+K^0$ and $B^+ \to \pi^0K^+$ [14]. Their amplitudes can be written in terms of diagrams as

$$
\begin{aligned}
A(B^+ \to \pi^+\pi^0) &= -\frac{1}{\sqrt{2}}(T + C) \ , \\
A(B^+ \to \pi^+K^0) &= P' \ , \\
A(B^+ \to \pi^0K^+) &= -\frac{1}{\sqrt{2}}(T' + C' + P') \ .
\end{aligned}
\tag{13}
$$

Thus, we have (what else?) triangle relations involving these decays (and their charge conjugates):

$$
\begin{aligned}
\sqrt{2}A(B^+ \to \pi^0K^+) + A(B^+ \to \pi^+K^0) &= r_u\sqrt{2}A(B^+ \to \pi^+\pi^0) \ , \\
\sqrt{2}A(B^- \to \pi^0K^-) + A(B^- \to \pi^-\overline{K^0}) &= r_u\sqrt{2}A(B^- \to \pi^-\pi^0) \ .
\end{aligned}
\tag{14}
$$

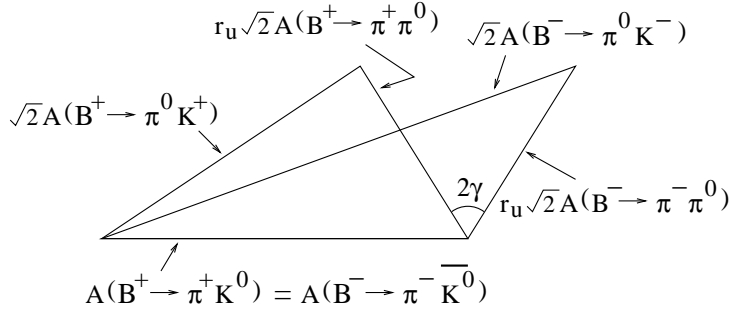

Figure 9: $SU(3)$ triangles involving decays of charged $B$'s, used to measure $\gamma$.

Let's look more closely at these triangle relations. First, although there are two diagrams contributing to $B^+ \to \pi^+\pi^0$, there is really only a single amplitude, since both $T$ and $C$ have the same weak phase. In isospin language, the two $\pi$'s are in a pure $I = 2$ state. Thus,

$$\sqrt{2}A(B^+ \to \pi^+\pi^0) = a_2 e^{i\gamma} e^{i\delta_2} \ . \tag{15}$$

Therefore $A(B^+ \to \pi^+\pi^0)$ and $A(B^- \to \pi^-\pi^0)$ have the same magnitude and have a relative phase $2\gamma$. Second, the decay $B^+ \to \pi^+K^0$ is pure penguin, i.e.

$$A(B^+ \to \pi^+K^0) = a_{P'} e^{i\phi_{P'}} e^{i\delta_P} \ . \tag{16}$$

However, $\phi_{P'} = \text{Arg}(V_{tb}^* V_{ts}) = \pi$, so we have $A(B^+ \to \pi^+K^0) = A(B^- \to \pi^-K^0)$. Thus, the two triangles share a base, and have a second side of equal length. As to the third side, the decay $B^+ \to \pi^0K^+$ has two contributions — $(T' + C')$ and $P'$ — which have different weak phases. Thus they can interfere, giving rise to direct CP violation. We therefore expect that, in general, $|A(B^+ \to \pi^0K^+)| \neq |A(B^- \to \pi^0K^-)|$.

By measuring the above rates, the two triangles can be constructed. As shown in Fig. 9, the construction can be used to extract the CKM angle $\gamma$, up to a discrete ambiguity due to the possible reflection of one of the triangles. Note the similarity to the $B \to DK$ construction described in the previous section – since only charged $B$'s are involved, neither tagging nor time dependence is needed. The advantage of this method compared to $B \to DK$ is that all branching ratios are of $O(10^{-5})$, so that in this case the triangles are not expected to be too thin.

$SU(3)$-breaking effects can be taken into account. Assuming factorization, the decays $B \to \pi\pi$ should include a factor $f_\pi$, while $B \to \pi K$ should have $f_K$ [12]. Thus, the factor $r_u$ which appears in the above construction should be multiplied by $f_K/f_\pi \approx 1.2$.

The above example showed how to obtain one of the CKM angles using $SU(3)$ relations. However, there is much more [15]! As mentioned above, the amplitudes for $B \to \pi\pi$, $\pi K$ and $KK$ can all be written in terms of the four diagrams $T$, $C$, $P$ and $P'$. It is convenient to write explicit expressions for these:

$$\begin{aligned} T &= a_T e^{i\gamma} e^{i\delta_T} \ , \\ C &= a_C e^{i\gamma} e^{i\delta_C} \ , \end{aligned}$$

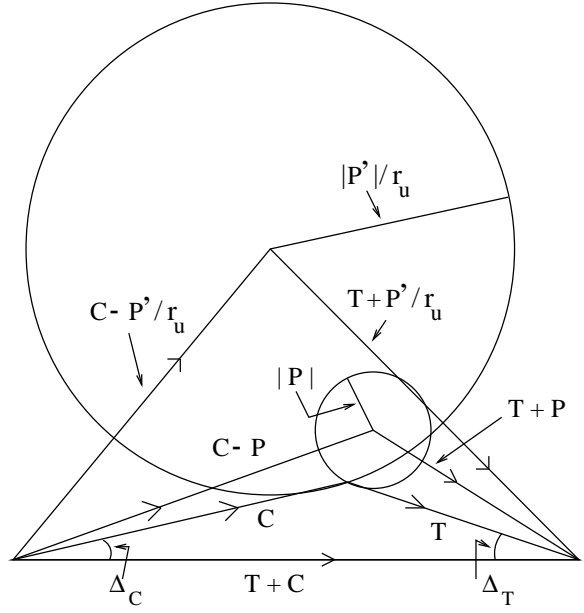

Figure 10: Two-triangle construction based on Eqs. 18 and 19 used to obtain weak and strong phases.

$$
\begin{aligned}
P &= a_P e^{-i\beta} e^{i\delta_P} \ , \\
P' &= -a_{P'} e^{i\delta_P} \ ,
\end{aligned}
\tag{17}
$$

Note that $|P|$ is measured in $B^+ \to K^+ \overline{K^0}$ or $B^0 \to K^0 \overline{K^0}$, while $|P'|$ can be obtained from $B^+ \to \pi^+ K^0$. Now consider the isospin triangle relation in Eq. 6. In $SU(3)$ diagram language, this can be written

$$
\begin{array}{ccccc}
A(B_d^0 \to \pi^+\pi^-) + \sqrt{2}A(B_d^0 \to \pi^0\pi^0) &=& \sqrt{2}A(B^+ \to \pi^+\pi^0) \\
(T+P) & + & (C-P) &=& (T+C) \ .
\end{array}
\tag{18}
$$

There is another triangle relation involving $\pi\pi$ and $\pi K$ final states. It is

$$
\begin{array}{ccccc}
\dfrac{1}{r_u}A(B_d^0 \to \pi^- K^+) + \dfrac{1}{r_u}\sqrt{2}A(B_d^0 \to \pi^0 K^0) &=& \sqrt{2}A(B^+ \to \pi^+\pi^0) \\
\left(T+\dfrac{P'}{r_u}\right) & + & \left(C-\dfrac{P'}{r_u}\right) &=& (T+C) \ .
\end{array}
\tag{19}
$$

Now, although these two triangles share a common base $[(T+C)]$, they do not appear to have anything else in common. However, they do – they share a common 'sub-triangle,' defined by

$$
T+C = (T+C) \ ,
\tag{20}
$$

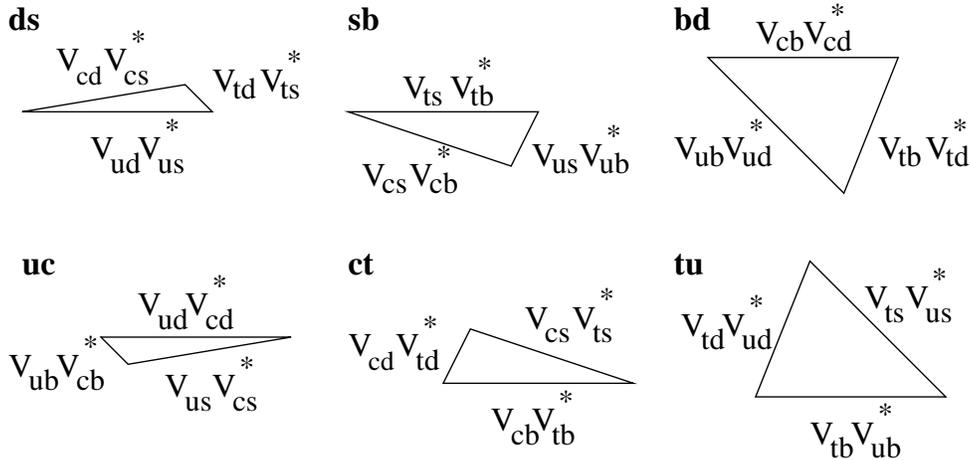

Figure 11: The six unitarity triangles. To the left of each triangle is indicated the pair of columns, or of rows, whose orthogonality this closed triangle expresses.

in which the combination $(T + C)$ is defined in Eq. 15. This sub-triangle is determined, up to discrete ambiguities, by the measurements of the 5 rates which make up the two triangles above, along with 2 rates which measure $|P|$ and $|P'|$. The construction which does this is shown in Fig. 10 (in this figure, $\Delta_i \equiv \delta_i - \delta_2$). The intersection of the two circles gives the apex of the sub-triangle. Note that this can all be done without tagging or time dependence.

Having obtained the sub-triangle, one extracts $a_T$, $a_C$, $\Delta_T$, $\Delta_C$, $\Delta_P + \alpha$ and $\Delta_P - \gamma$. These latter two quantities can be combined to yield the CKM angle $\beta$. Note that it is not even necessary to measure the CP-conjugate rates to obtain $\beta$. However, if one also measures the CP-conjugate rates, then one can obtain $\alpha$, $\gamma$ and $\Delta_P$ separately. In other words, using this technique, *it is possible to obtain the weak CKM phases, the strong final-state phase differences, and the sizes of the individual diagrams!* And, as before, $SU(3)$ breaking can be taken into account by letting $r_u \rightarrow r_u f_K / f_\pi$.

There are other such two-triangle constructions [15], which provide additional ways of measuring the same quantities. This redundancy can be used to further test the $SU(3)$ symmetry. Again, in all of these analyses, neither tagging nor time dependence are needed, although time-dependent CP asymmetries will yield further information. Perhaps both symmetric and asymmetric $e^+e^-$ colliders will be competing to map out the unitarity triangle.

### An exact treatment of the CKM matrix

Up to now, all our efforts have been concentrated on finding ways of cleanly measuring the angles of the unitarity triangle, which is based on an approximate parametrization of the CKM matrix. However, if one uses an *exact* parametrization of the CKM matrix, there are 6 unitarity triangles (Fig. 11), corresponding to the orthogonality of the 3 rows and 3 columns. (The unitarity triangle of Fig. 1 is just the *bd* triangle of Fig. 11). It turns out to be quite interesting to study these 6 triangles. Due to space limitations, I will only sketch out some of the results, but I refer the reader to Ref. [16] for the details.

There are 18 interior angles in these 6 unitarity triangles. It is fairly straightforward to show that at most 4 of these are independent. This is not surprising – the CKM matrix can be parametrized by 4 parameters, typically taken to be 3 angles and a complex phase. Thus it is impossible for there to be more than 4 independent interior angles in the unitarity triangles. What is more surprising, perhaps, is that these 4 interior angles *are*, in fact, independent. In other words, they form a parametrization of the CKM matrix.

A convenient choice for the 4 independent angles is the following: two of the angles in the $bd$ triangle (e.g. $\alpha$ and $\beta$), the small angle ($\epsilon$) in the $sb$ triangle, and the tiny angle ($\epsilon'$) in the $ds$ triangle. From our knowledge of the sizes of the CKM matrix elements, we know that $\alpha$ and $\beta$ can be big, $\epsilon \lesssim 0.05 \ [O(\lambda^2)]$, and $\epsilon' \lesssim 0.003 \ [O(\lambda^4)]$. What is interesting about this is that all 4 angles can, in principle, be measured through CP violation in the $B$ system! The angles $\alpha$ and $\beta$ can be obtained through rate asymmetries in $\overset{(-)}{B_d} \to \pi^+\pi^-$ and $\overset{(-)}{B_d} \to \Psi K_s$, as discussed previously. The angle $\epsilon$ can be measured in $\overset{(-)}{B_s} \to \Psi\phi$ (recall that this is the fourth class of CP asymmetries, expected to vanish in the Wolfenstein parametrization.) Finally, the tiny angle $\epsilon'$ can, in principle, be measured by comparing the value of $\gamma$ as obtained via CP asymmetries in the decays $\overset{(-)}{B_s} \to D_s^+ K^-, \ D_s^- K^+$ and $B^\pm \to D_{CP}^0 K^\pm$ with that as deduced from the values of $\alpha$ and $\beta$ – in the Wolfenstein approximation all measure $\gamma$, but in an exact treatment they differ by different functions of $\epsilon$ and $\epsilon'$. Thus, in principle, by using measurements of CP violation in the $B$ system alone, one can reconstruct the entire CKM matrix!

Of course, in practice, we will never be able to get at the tiny angle $\epsilon'$. It may be possible, however, to eventually measure $\epsilon$. If so, then this will give us another way to measure such quantities as $|V_{ub}/V_{cb}|$ and $|V_{td}/V_{ts}|$. Specifically,

$$\left|\frac{V_{ub}}{V_{cb}}\right|^2 \simeq \frac{\sin\beta\sin\epsilon}{\sin\alpha\sin\gamma} \ , \quad \left|\frac{V_{td}}{V_{ts}}\right|^2 \simeq \frac{\sin\gamma\sin\epsilon}{\sin\alpha\sin\beta} \ . \tag{21}$$

The advantage of this method is that $|V_{ub}/V_{cb}|$ and $|V_{td}/V_{ts}|$ can be obtained cleanly, i.e. with no theoretical uncertainty.

## CONCLUDING REMARKS

CP violation in the $B$ system can be used to test one of the few remaining untested areas of the standard model, the CKM matrix. There are *many* ways to cleanly extract CKM phase information from CP-violating rate asymmetries in $B$ decays, and these measurements will be made in the near future. Will this give us our first glimpse of physics beyond the standard model?

## ACKNOWLEDGEMENTS

I would like to thank R. Aleksan, M. Gronau, O. Hernández, B. Kayser, R.D. Peccei and J.L. Rosner for collaborations and helpful conversations. This work was supported in part by the N.S.E.R.C. of Canada and les Fonds F.C.A.R. du Québec.